%% file: main.tex
\documentclass[reprint,
superscriptaddress,
amsmath,amssymb,
aps,
prb,
floatfix]{revtex4-2}

\input{packages}
\input{editing}

\input{glossary.tex}

\input{acronyms.tex}
\input{commands.tex}



\begin{document}

    \title{Magnetic properties of Cr$_8$ and V$_8$ molecular rings from ab initio calculations}
        
    \author{Elia Stocco}
    \affiliation{Department of Physics, University of Pavia, Via A. Bassi 6, I-27100 Pavia, Italy}
    
    \author{Maria Barbara Maccioni}
    \affiliation{Department of Physics, University of Pavia, Via A. Bassi 6, I-27100 Pavia, Italy}
    
    \author{Andrea Floris}\email[e-mail:]{
    afloris@lincoln.ac.uk} 
    \affiliation{Department of Chemistry, School of Natural Sciences, University of Lincoln, Brayford Pool, Lincoln LN6 7TS, United Kingdom}
    
    \author{Matteo Cococcioni}
    \email[e-mail:]{matteo.cococcioni@unipv.it}
    \affiliation{Department of Physics, University of Pavia, Via A. Bassi 6, I-27100 Pavia, Italy}
    
    \date{\today}

        \input{abstract}

    \maketitle

    \color{black}

    \input{1-introduction}
    \input{2-complexes}
    \input{3-theory}

    \input{4-computational}

    \input{5-results}
    \input{6-discussion}
    \input{7-conclusions}
    \input{acknowledgements}

        \bibliographystyle{unsrt}
\bibliography{bibliography}

\end{document}

%% file: packages.tex
\usepackage[utf8]{inputenc}
\DeclareUnicodeCharacter{0301}{\'{}}

\usepackage{bm}
\usepackage[acronym,toc,automake=immediate,nonumberlist,nopostdot,nogroupskip]{glossaries}

\usepackage[hidelinks]{hyperref}
\hypersetup{
    colorlinks=true,
    breaklinks=true,
    linkcolor=blue,
    filecolor=magenta,      
    urlcolor=blue,
    pdftitle={Magnetic properties of Cr8 and V8 molecular rings from ab initio calculations},
    pdfpagemode=FullScreen,
}

\usepackage{mathtools,amssymb,amscd,amsthm,amsmath,graphicx,color,bbm,bm,amsbsy,amsfonts}
\usepackage{braket}

\usepackage{array}
\newcolumntype{R}[1]{>{\raggedleft\arraybackslash}p{#1}}
\newcolumntype{L}[1]{>{\raggedright\arraybackslash}p{#1}}
\newcolumntype{C}[1]{>{\centering\arraybackslash}p{#1}}

\usepackage{xspace}
\usepackage{multirow}

%% file: editing.tex
\newboolean{preview}   
\setboolean{preview}{false} 
\ifthenelse{\boolean{preview}}{%
\newcommand{\editor}[2]{%
  \expandafter\newcommand\csname #1note\endcsname[1]{\ignorespaces}%
  \expandafter\newcommand\csname #1cancel\endcsname[1]{\ignorespaces}%
  \expandafter\newcommand\csname #1change\endcsname[2]{##2}%
  \newenvironment{#1text}{\color{#2}}{\color{black}}
  \expandafter\newcommand\csname #1add\endcsname[1]{##1}
}%
}{%
\newcommand{\editor}[2]{%
  \expandafter\newcommand\csname #1note\endcsname[1]{%
    \textcolor{#2}{(\textbf{#1:} ##1)}}%
  \expandafter\newcommand\csname #1cancel\endcsname[1]{%
    \textcolor{#2}{\sout{##1}}}%
  \expandafter\newcommand\csname #1change\endcsname[2]{%
    \textcolor{#2}{\sout{##1} ##2}}%
  \newenvironment{#1text}{\color{#2}}{\color{black}}
  \expandafter\newcommand\csname #1add\endcsname[1]{%
    \textcolor{#2}{##1}}
}%
}

\definecolor{red}{rgb}{1.00,0.00,0.00}
\definecolor{green}{rgb}{0.00,1.00,0.00}
\definecolor{blue}{rgb}{0.00,0.00,1.00} 
\definecolor{orange}{rgb}{1.00,0.50,0.00}
\definecolor{ORANGE}{rgb}{1.00,0.50,0.00}
\definecolor{magenta}{rgb}{1.00,0.00,1.00}
\definecolor{cyan}{rgb}{0.00,1.00,1.00}
\definecolor{brown}{rgb}{0.4,0.2,0.00}
\definecolor{deepsky}{rgb}{0.0,0.75,1.00}
\definecolor{gray}{rgb}{0.50,0.50,0.50}
\definecolor{navy}{rgb}{0.137,0.137,0.557}
\definecolor{burntorange}{rgb}{0.8, 0.33, 0.0}
\definecolor{asparagus}{rgb}{0.53, 0.66, 0.42}
\definecolor{emerald}{rgb}{0, 0.605, 0.465}
\definecolor{amethyst}{rgb}{0.6, 0.4, 0.8}
\definecolor{purple}{rgb}{0.502, 0, 0.502}
\definecolor{lava}{rgb}{0.81, 0.06, 0.13}
\definecolor{tk}{rgb}{0.5, 0.00, 0.8}
\definecolor{beaver}{rgb}{0.62, 0.51, 0.44}
\definecolor{yellow}{rgb}{0.7, 0.7, 0.0}
\definecolor{seagreen}{rgb}{0.18,0.55,0.34}
\definecolor{oxford}{rgb}{0.0,0.129,0.278}
\definecolor{lightgray}{rgb}{.9,.9,.9}
\definecolor{lightblue}{rgb}{0.0,0.0,0.8}
\definecolor{fuchsia}{rgb}{1.0,0.47,1.0}

\editor{AF}{black}
\editor{MC}{black}
\editor{ES}{black}
\editor{BM}{black}

%% file: glossary.tex
\makeglossaries  

\newglossaryentry{HK}{
    name=Hohenberg–Kohn,
    description={The Hohenberg–Kohn theorem, which states that the energy of the ground electronic state is a unique functional of the electron density, provides the foundation for density functional theory.} 
}

\newglossaryentry{hei}{
    name=Heisenberg,
    description={Heisenberg} 
}

\newglossaryentry{nc}{
    name=non-collinear,
    description={non-collinear} 
}

\newglossaryentry{r}{
    name=Ref.,
    description={reference} ,
    plural={Refs.}
}

\newglossaryentry{dof}{
    name=degrees of freedom,
    description={degrees of freedom} 
}

\newglossaryentry{BO}{
    name=Born-Oppenheimer,
    description={Born-Oppenheimer} 
}

\newglossaryentry{pp}{
    name=pseudopotential,
    description={pseudopotential} 
}

\newglossaryentry{pw}{
    name=plane waves,
    description={plane waves} 
}

\newglossaryentry{PSlib}{
    name=PSlibrary,
    description={PSlibrary} 
}

\newglossaryentry{Co}{
    name=Coulomb,
    description={Coulomb} 
}

\newglossaryentry{Ha}{
    name=Hartree,
    description={Hartree} 
}

\newglossaryentry{KS}{
    name=Kohn-Sham,
    description={Kohn-Sham.} 
}

\newglossaryentry{cdf}{
    name={charge density},
    description={charge density} ,
    plural={charge densities},
    longplural={charge densities},
    shortplural={charge densities}
} 

\newglossaryentry{sppol}{
    name={spin polarization},
    description={pin polarization} 
    plural={pin polarization}
    longplural={pin polarization}
} 

\newglossaryentry{wf}{
    name=wave function,
    description={wave function} 
} 

\newglossaryentry{ham}{
    name=Hamiltonian,
    description={Hamiltonian} 
}

\newglossaryentry{AFM}{
	name={antiferromagnet}, 
	description={antiferromagnet}
}

\newglossaryentry{FM}{
	name={ferromagnet}, 
	description={ferromagnet}
}

\newglossaryentry{afm}{
	name={antiferromagnetic}, 
	description={antiferromagnetic}
}

\newglossaryentry{fm}{
	name={ferromagnetic}, 
	description={ferromagnetic}
}

\newglossaryentry{RHF} {
    name={RHF},
    description={Restricted Hartree Fock},
    first={Restricted Hartree Fock (RHF)},
    text={RHF}
}

\newglossaryentry{UHF} {
    name={URH},
    description={Unrestricted Hartree Fock},
    first={Unrestricted Hartree Fock (URH)},
    text={URH}
}

\newglossaryentry{ma}{
	name={MA}, 
	description={Magnetic Anisotropy}, 
	first={Magnetic Anisotropy (MA)}, 
	text={MA}
}

\newglossaryentry{xcd}{
	name={XCrySDen}, 
	description={XCrySDen (\textit{(X-Window) CRystalline Structures and DENsities},\url{http://www.xcrysden.org/XCrySDen.html})}, 
	first={XCrySDen (\textit{(X-Window) CRYstalline Structures and DENsities})}, 
	text={XCrySDen}
}

\newglossaryentry{VESTA}{
	name={VESTA}, 
	description={VESTA (\textit{Visualization for Electronic and STructural Analysis},\url{https://jp-minerals.org/vesta/en/})}, 
	first={VESTA (\textit{Visualization for Electronic and STructural Analysis})}, 
	text={VESTA}
}

\newglossaryentry{QE}{
	name={Quantum ESPRESSO}, 
	description={Quantum ESPRESSO}, 
	first={Quantum ESPRESSO (\textit{opEn Source Package for Research in Electronic Structure, Simulation, and Optimization)}}, 
	text={Quantum ESPRESSO}
}

\newglossaryentry{FLL}{
	name={FLL}, 
	description={Fully Localized Limit}, 
	first={Fully Localized Limit (FLL)}, 
	text={FLL}
}

\newglossaryentry{gs}{
    name=ground state,
    description={Ground-state.} 
    first={ground-state}, 
}

\newglossaryentry{sie}{
	name={self-interaction error}, 
	description={self-interaction error}, 
	first={self-interaction error (SIE)}, 
	text={SIE}
}

\newglossaryentry{sic}{
	name={self-interaction correction}, 
	description={self-interaction correction}, 
	first={self-interaction correction (SIC)}, 
	text={SIC}
}

\newglossaryentry{raa}{
    name=\textit{reductio ad absurdum},
    description={In logic, \textit{reductio ad absurdum} is the form of argument that attempts to establish a claim by showing that the opposite scenario would lead to absurdity or contradiction.} 
}

\newglossaryentry{SH}{
	name={spin Hamiltonian}, 
	description={spin Hamiltonian}
}

\newglossaryentry{evd}{
	name={E-V-density}, 
	description={ensemble $v$-representable densities}, 
	first={ensemble $v$-representable densities (E-V-densities)}, 
	text={evd}
}

\newglossaryentry{CSi}{
    name=Cauchy–Schwarz inequality,
    description={The Cauchy–Schwarz inequality is considered one of the most important and widely used inequalities in mathematics.} 
}

\newglossaryentry{sc}{
    name=\text{Schr\"{o}dinger},
    description={Erwin Rudolf Josef Alexander \text{Schr\"{o}dinger}, was a Nobel Prize-winning Austrian-Irish physicist who developed a number of fundamental results in quantum theory: the \text{Schr\"{o}dinger} equation provides a way to calculate the wave function of a system and how it changes dynamically in time.} 
}

\newglossaryentry{in}{
    name=interaction,
    description={interaction} 
}

\newglossaryentry{lo}{
    name=\text{L{\"o}wdin},
    description={\text{L{\"o}wdin}} 
}

\newglossaryentry{algo}{
    name=algorithm,
    description={algorithm} 
}

\newglossaryentry{ai}{
    name=ab initio,
    description={ab initio} 
}

\newglossaryentry{cou}{
    name=Coulomb,
    description={Coulomb} 
}

\newglossaryentry{e}{
    name=electron,
    description={electron} 
}

\newglossaryentry{ec}{
    name=electronic,
    description={electronic} 
}

\newglossaryentry{con}{
    name=configuration,
    description={configuration} 
}

\newglossaryentry{spcon}{
    name=spin configuration,
    description={spin configuration} 
}

\newglossaryentry{HF}{
    name=Hellmann–Feynman,
    description={In quantum mechanics, the Hellmann–Feynman theorem relates the derivative of the total energy with respect to a parameter, to the expectation value of the derivative of the Hamiltonian with respect to that same parameter.} 
}

\newglossaryentry{hf}{
    name=Hartree-Fock,
    description={Hartree-Fock} 
}

\newglossaryentry{s}{
    name=susceptibility,
    description={susceptibility} ,
    plural={susceptibilities}
}

\newglossaryentry{ms}{
    name=magnetic susceptibility,
    description={magnetic susceptibility} 
}

\newglossaryentry{dms}{
    name=differential magnetic susceptibility,
    description={differential magnetic susceptibility} 
}

\newglossaryentry{oam}{
	name={orbital angular momentum}, 
	description={orbital angular momentum}
}


%% file: acronyms.tex
\newacronym{nn}{n.n.}{nearest neighbor}
\newacronym{DFT}{DFT}{Density Functional Theory}
\newacronym{NMR}{NMR}{Nuclear Magnetic Resonance}
\newacronym{TM}{TM}{transition metal}
\newacronym{EPR}{EPR}{Electron Paramagnetic Resonance}
\newacronym{INS}{INS}{Inelastic Neutron Scattering}
\newacronym{bfgs}{BFGS}{Broyden-Fletcher-Goldfarb-Shanno}
\newacronym{bs}{BS-DFT}{Broken-Symmetry DFT}
\newacronym{gka}{GKA}{Goodenough-Kanamori-Anderson}
\newacronym{SQUID}{SQUID}{Superconducting QUantum Interference Device}
\newacronym{DM}{DM}{Dzyaloshinskii–Moriya}
\newacronym{uspp}{USPP}{Ultra-Soft Pseudo-Potentials}
\newacronym{LR}{LR}{Linear Response}
\newacronym{DFPT}{DFPT}{Density Functional Perturbation Theory}
\newacronym{LDA}{LDA}{Local Density Approximation}
\newacronym{zfs}{ZFS}{Zero Field Splitting}
\newacronym{LSDA}{LSDA}{Local Spin Density Approximation}
\newacronym{GGA}{GGA}{Generalized Gradient Approximation}
\newacronym{PBE}{PBE}{Perdew-Burke-Ernzherof}
\newacronym{xc}{xc}{exchange and correlation}
\newacronym{pdos}{PDOS}{Projected Density Of States}
\newacronym{MNM}{MNM}{Molecular Nano-Magnet}
\newacronym{soc}{SOC}{spin-orbit coupling}
\newacronym{SMM}{SMM}{Single-Molecule Magnets}
\newacronym{pbc}{p.b.c.}{periodic boundary conditions}
\newacronym{RMSE}{RMSE}{Root Mean Square Error}
\newacronym{cf}{CF}{crystal field}

%% file: commands.tex
\newcommand\dep[1]{\ensuremath{\left(#1\right)}}

\def\restrict#1{\raise-.5ex\hbox{\ensuremath|}_{#1}}

\newcommand{\pos}[1]{\mathbf{#1}}

\newcommand{\mypar}[1]{}

\newcommand{\Jone}{\ensuremath{J^{\dep{1}}}\xspace} 
\newcommand{\Jtwo}{\ensuremath{J^{\dep{2}}}\xspace} 

%% file: abstract.tex
\begin{abstract}
    
\noindent%

Molecular nanomagnets are systems with a vast phenomenology and are very  promising for a variety of technological applications, most notably spintronics and quantum information. 
Their low-energy spectrum and magnetic properties can be modeled using effective spin Hamiltonians, once the  exchange coupling parameters   between the localized magnetic moments are determined. 
In this work we employ density functional theory (DFT)  to compute the exchange parameters between the atomic spins for two representative ring-shaped molecules containing eight transition-metal magnetic ions: Cr$_8$ and V$_8$. 
Considering a set of properly chosen spin configurations and mapping their DFT energies on the corresponding expressions from a Heisenberg Hamiltonian,
we compute the exchange couplings between magnetic ions which are first, second and further neighbors on the rings.
In spite of their chemical and structural similarities the two systems exhibit very different ground states:
antiferromagnetic for Cr$_8$, ferromagnetic for V$_8$, which also features non-negligible couplings  between second nearest neighbors. A rationalization of these results is proposed that is based on a multi-band Hubbard model with less-than-half filled shells on magnetic ions. 

\glsresetall

\end{abstract}

%% file: 1-introduction.tex
\section{Introduction}
\label{sec:intro}

\newcommand{\niceheight}{0.24\textheight}

\begin{figure*}[!htb]
    \hspace{3mm}
    \begin{minipage}[t]{0.45\textwidth}
        \centering
        \includegraphics[height=0.24\textheight]{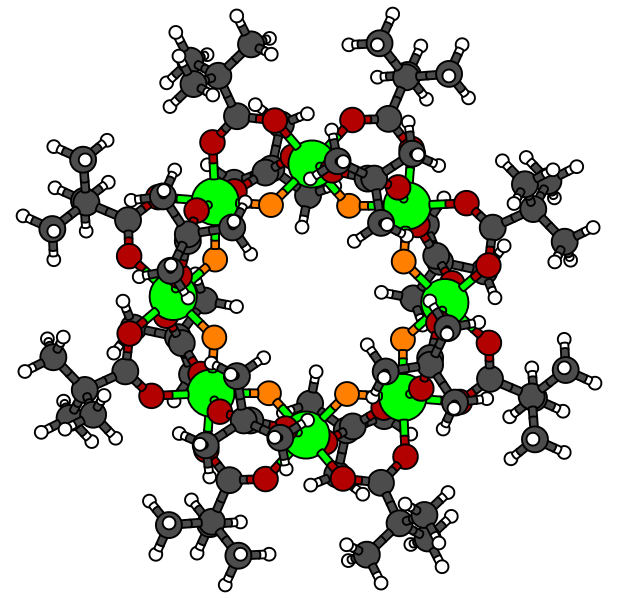}
        \caption{Cr$_8$/V$_8$ molecular structure. Colors code: Cr/V green, F orange, O red, C grey, H white.}
        \label{fig:geometries-molecule}
    \end{minipage}%
    \hspace{5mm} 
    \begin{minipage}[t]{0.45\textwidth}
        \centering
        \includegraphics[height=0.24\textheight]{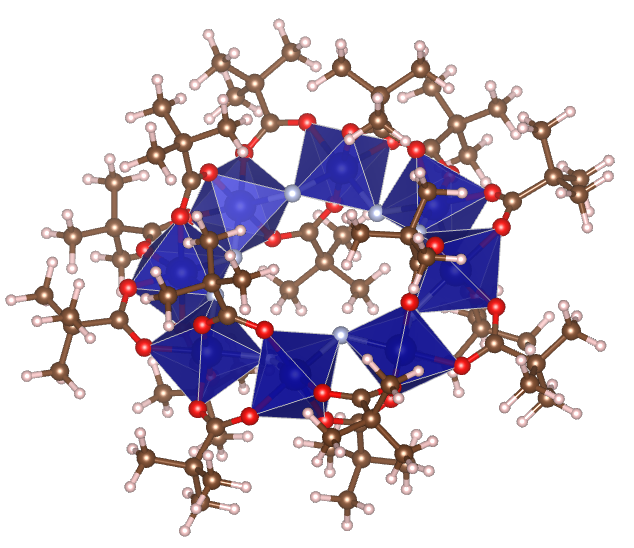}
        \caption{3D molecular structure highlighting the anion octahedra around each \gls{TM} ion.}
        \label{fig:octahedra}
    \end{minipage}
\end{figure*}

Molecular nanomagnets (MNM) are attracting a vast interest in the scientific community, due to their huge  potential for 
technological applications \cite{Garlatti}, including spintronics \cite{walker2016molecular, Sun2015, Jungwirth2018, Smejkal2018, Jungwirth2016}, magneto-refrigeration \cite{spichkin2001magnetic, affronte2004engineering, schnack2007enhanced, evangelisti2010recipes}, miniaturisation of data-storage media \cite{Mannini2009}, quantum sensing \cite{troiani19} and quantum-information 
\cite{troiani11,Leuenberger2001, PhysRevLett.98.057201, PhysRevLett.108.107204, timco2009engineering, PhysRevLett.94.190501, PhysRevLett.94.207208, B615543J, aec39e306fd34bdabf465a5edee31b20, PhysRevLett.104.037203, PhysRevLett.107.230502,Carretta2024,Coronado,AtzoriSessoli}. 
The reduced size of these systems and, in many cases, the availability of experimental data for comparison, make them ideal  
to even investigate  fundamental questions such as the different behavior of integer and half-integer spin systems \cite{HALDANE1983464,Kennedy1992,PhysRevB.45.304,HaldaneQuantumSpinChains,YAMASHITA2000347,DARRIET1993409,PhysRevLett.56.371}, the influence of geometrical curvature  on the spin order and  magnetic interactions \cite{gentile13,gentile15}, the origin of generalized spin-orbit couplings 
 \cite{cardias2020dzyaloshinskii}, \and the relation between spin and transport properties \cite{PhysRevLett.102.246801}.
Here, we consider a specific class of MNM: ring-shaped molecules with eight magnetic centers \AFadd{consisting of transition-metal (TM) ions carrying localized spins,} 
occupied by either Cr or V, known as Cr$_8$ and V$_8$, respectively.
Due to their modular and stable  structure \AFadd{composed of} TM-centered anion octahedra glued together  by organic ligands, and the interchangeable TM ions, they offer a  platform to  control a variety of magnetic  behaviors, which could be engineered for target applications.
In fact, these systems have been studied for over two decades, both experimentally \cite{affronte03,whinpenny,Lascialfari,Waldmann03,van2002magnetic} and theoretically \cite{bellini2010density,Kronik,chiesa16}, with particular attention to Cr-based rings,  due to their potential for quantum information technologies \cite{chiesa23,chiesa14,chiesa20,ghirri17,timco09,timco16,candini10}.

Cr$_8$ is the prototype
 of this class of systems and has been considered both in its pristine and doped compositions   
with one  Cr atom substituted by other TMs to achieve a finite total spin   or other interesting magnetic properties \cite{Larsen,Kronik,slusarski2011electronic,bellini2006density,Bellini2008,PhysRevLett.110.157204,PhysRevB.67.094405,Waldmann03,baker2012spin,PhysRevB.77.014402,van2002magnetic,bellini2010density}.
%
Cr$_8$ features an antiferromagnetic (AFM) order across the ring, with the Cr half integer spins ($S=3/2$) aligned perpendicularly to the molecular plane,  
in a collinear fashion. 
This order is quite common among eight-center rings based on 3d  magnetic centers.

In contrast, octanuclear  rings based on V,
like V$_8$, have been studied  sporadically in literature \cite{sorolla19,laye2003solvothermal} and never with the specific structure/composition we investigate here, due to its instability. The only studies we are aware of on V-based rings are on doped magnetic wheels of the form V$_7$X, where X is a TM different from V  \cite{Lascialfari}). Thus, to the best of our knowledge, this work is the first  to address the behavior of V$_8$. 
Due to the 3+ oxidation state of the TM ions, V$_8$ features
an integer $S=1$ spin on each magnetic center,  consistent with the V ion having one less 3d electron than Cr. However, at variance with similar octanuclear rings based on Cr or other 3d ions, we find V$_8$  to exhibit a peculiar ferromagnetic (FM) behaviour, also characterized by 
non-negligible AFM couplings between second nearest neighbor magnetic ions. 
  
In this work, we present a first computational study of  V$_8$  aimed at investigating, in comparison with the well-known Cr$_8$, its peculiar ground state magnetic properties. Density functional theory \cite{PhysRev.136.B864,KS65} calculations are used to determine  the electronic \gls{gs} of both systems and compare the energy of various spin configurations.  
%
%
Precise values of the exchange couplings between the TM ions are obtained from DFT, upon fixing the \AFadd{sign 
of the spin} on each magnetic site, while their magnitude remains unchanged.
%
%

The paper is organized as follows: in Section \ref{sec:structures} we introduce the Cr$_8$ and V$_8$ molecular structures. 
In Section \ref{sec:theory} we briefly outline the theory of spin Hamiltonians and their integration with DFT calculations. 
The results of our magnetic couplings calculations  
are presented in Section \ref{sec:res}, 
while in Section \ref{sec:disc}
we discuss our main findings, giving a rationale of the various approximations used.  
Section \ref{sec:conclusions} concludes and summarizes  
the work.

%% file: 2-complexes.tex
\section{Cr$_8$ and V$_8$ molecular complexes}\label{sec:structures}



The Cr$_8$ and V$_8$ rings  have chemical formula M$_8$F$_8$Piv$_{16}$, where Piv is the pivalic group (CH$_3$)$_3$CCO$_2$ and M$\equiv$Cr/V. 
%
Aside the chemical identity of the \gls{TM} ions \AFadd{and} minimal changes in the interatomic distances and bond angles, the Cr$_8$ and V$_8$  structures are essentially identical (Fig.~\ref{fig:geometries-molecule}). 
The \gls{TM} ions are 
coordinated by six anions (two F and four O) forming an 
octahedron around each of them, as shown in Fig.~\ref{fig:octahedra}.
The octahedra are somewhat distorted and tilted  \AFadd{to fit into a ring-shaped geometry, and they share their inner corners \AFadd{towards} the center of the molecule, which are occupied by F anions.}
Externally, instead, they are
connected via two carboxylate (OCO) bridges, part of the dangling pivalic groups. One carboxylate lies on the ring's plane, while the other is nearly perpendicular to it.
The perpendicular carboxylates show alternate orientation, one pointing upward and the next downward resulting in a $D_4$ 
symmetry of the molecular structure \cite{walsh}.  
%
The stoichiometry of these molecules imposes a nominal $3+$ oxidation state to each \gls{TM} ion, with the V$^{3+}$ and Cr$^{3+}$ electronic configurations being [Ar]$3d^2$ and [Ar]$3d^3$, respectively. According to Hund's first rule, they correspond to ground states with \AFadd{integer} spin $S=1$  for V$^{3+}$ and \AFadd{semi-integer} $S=3/2$  for Cr$^{3+}$.   
Due to the octahedral coordination shell, the reference structure of the $d$ states spectrum is the one produced by a cubic crystal field, with \AFadd{the two- and three-fold subgroups 
 $e_g$ and $t_{2g}$.} 
This is actually only approximately true in the considered systems because the distorted TM-O/F octahedra 
have lost their cubic symmetry,  further lowering these subgroups degeneracy.
However, since crystal field interactions are generally weaker, \AFadd{albeit comparable to} the Hund's exchange, this $d$ states structuring does not affect the total spin of the central cation and \gls{TM} species are in a high-spin configuration.  
This situation can change substantially in presence of significant spin-orbit couplings as discussed, for example, in Ref. \cite{garlatti18}.

As mentioned earlier, Cr$_8$ is a well-known system, which has been synthesized and studied  thoroughly (even in presence of TM substitutional doping) both experimentally 
\cite{van2002magnetic, baker2012spin, PhysRevB.77.014402, PhysRevB.67.094405, Waldmann03}, 
and theoretically \cite{Kronik, bellini2010density, slusarski2011electronic, bellini2006density, Bellini2008,PhysRevLett.110.157204,PhysRevLett.110.157204} via DFT with hybrid functionals, Hubbard corrections, or in conjunction with many-body models.

Unlike Cr$_8$, the specific V$_8$ ring we consider in this work  has never been synthesized, due to parasitic oxidation processes  
\cite{Lascialfari}. Therefore, a direct comparison of our results with experiments is currently not feasible.  
To the best of our knowledge, the only two homometallic rings containing eight  V magnetic centers ever synthesized and characterized experimentally  
are reported in Refs. \cite{sorolla19} and \cite{laye2003solvothermal}. However, compared to our V$_8$, they feature a different organic scaffold around the TM centers, and, quite interestingly, report different ground-state magnetic configurations, \AFadd{FM and AFM, respectively}. On the other hand,
two heterometallic V-based rings with the same  structure as our V$_8$ (see Fig.~\ref{fig:geometries-molecule}) have been synthesized, with single Ni and Zn impurities  (V$_7$Ni and V$_7$Zn, respectively). Their experimental characterization shows an  AFM ground state \cite{walsh, Lascialfari, adelnia} which has led authors to expect a similar behavior also for V$_8$. A comparison of these systems with the ones studied here is left for future work.

%% file: 3-theory.tex
\section{Theory and computation}
\label{sec:theory}
In this section we present the theoretical framework and the technical details of our calculations. We first discuss the spin Hamiltonian considered in this work  and then outline the methodology used to compute the exchange interactions between the magnetic centers along the rings from ab initio calculations. 


\subsection{Spin Hamiltonian}
\label{sec:spinh}
 The spin Hamiltonian $\hat H_{ex}$ used in this work is of the Heisenberg type, with magnetic exchange couplings
 between first- and second-nearest-neighbor spins, while interactions between more distant neighbors turn out to be negligible. $\hat H_{ex}$ accounts for the coupling between localized moments through their scalar product:
        \begin{equation}
            \hat H_{ex} = J^{(1)} \sum_i \hat{\mathbf S}_i \cdot \hat{\mathbf S}_{i+1} +
            J^{(2)} \sum_i \hat{\mathbf S}_i \cdot \hat{\mathbf S}_{i+2}.
            \label{heis1}
        \end{equation}
        
\begin{figure}
            \centering
         \includegraphics[scale=0.6]{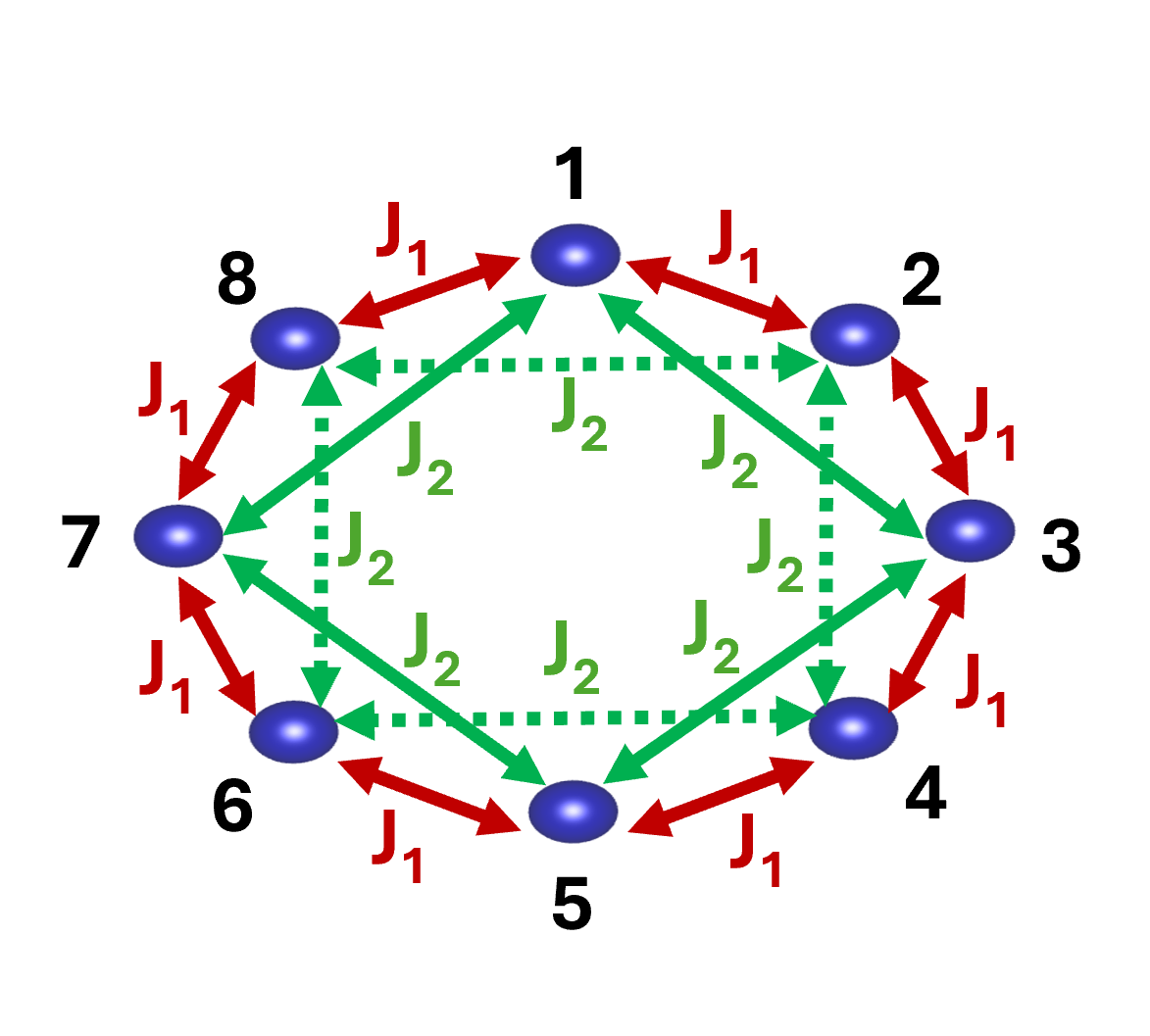}
            \caption{Exchange coupling parameters \Jone and \Jtwo   between magnetic ions.}
            \label{fig:j-config}
        \end{figure}

        Fig.~\ref{fig:j-config} offers a visual representation of the magnetic couplings between the TM centers along the molecule.

\subsection{Exchange interactions}\label{sec:CD}
        
        The magnetic couplings present in the spin Hamiltonian (Eq. \ref{heis1}) will be obtained from ab initio calculations based on DFT,  \cite{PhysRev.136.B864} in the \gls{KS} formulation \cite{KS65}. While the  full characterization of MNMs magnetic properties   would require  wavefunction-based quantum chemistry approaches, these methods are largely impractical here, due to the large size of our systems. DFT 
        represents an optimal compromise between accuracy and computational costs, 
        allowing direct access to the systems \gls{gs} energy  and selected perturbations around it.

        The magnetic couplings 
        are evaluated from ab initio calculations, by fitting the DFT total energy of suitably chosen spin configurations with the predictions of the spin Hamiltonian.   
        Besides widely used local and semilocal approximations to the exchange-correlation (xc) functional \AFadd{such as} LDA~\cite{PhysRev.136.B864, KS65} and GGA~\cite{Perdew:1996}, we use  the popular DFT+U  \cite{Anisimov1991,Anisimov1995} and the extended DFT+U+V \cite{DFTUV,Himmetoglu} approaches, necessary to improve the $d$ states localization and achieve a more accurate representation of the magnetic moments and their interactions.

        Spin-resolved DFT calculations are performed  within the collinear framework,   
        where each electron is in an eigenstate of the $\sigma_z$ Pauli matrix and the electronic magnetization is a scalar function. In this simplified framework, all  magnetic moments align parallel to the same spin quantization axis, independent of the crystalline axes orientation (a non-collinear generalization  will be discussed in a future work). 
        %
        %
        %
        %
        
        Over the last decades several methods have been introduced to compute the effective exchange couplings from first principles,  
        including those based on Green's functions and linear-response theory \cite{katsnelson99,korotin15,he21}, and  
        the four-state method \cite{xiang11,xiang13,PhysRevB.102.014457}.
        
        Here we follow a somewhat simpler approach  based on 
        the broken-symmetry DFT  method (BS-DFT) ~\cite{vanWullen2009, kessler2013broken, WullenChristoph, Schmitt, Ruiz2003, Ruiz1999, Ruiz2005, schurkus2020theoretical, C2DT31662E}.  %
        %
        %
        More specifically, we adopt the \textit{projected} BS-DFT, 
        a method which relies on the same assumptions of  
        the four-states method \cite{xiang11,xiang13,PhysRevB.102.014457}. 
        \AFadd{This approach assumes that the exchange coupling constants
        do not depend significantly on the electronic states, whose energy is compared to compute them. Other, more sophisticated methods based on quantum-chemistry, suggest that this might not be true for all systems ~\cite{spinham1,spinham2,spinham3}. 
        While DFT-based calculations cannot be expected to capture the energetics of multi-configurational states, the Hubbard corrections we employ here should make the energy of single spin-configurations quite reliable. These states are the basis of}
        the \AFadd{BS-DFT} approach which consists in mapping DFT total energies of selected spin configurations 
        onto the expectation values of the spin Hamiltonian (Eq.~\ref{heis1}) on the same states, 
        \AFadd{i.\,e.} on single-determinants constructed from the appropriate eigenstates of $\hat S^z$ on each magnetic ion.  
        In practice, if we indicate these states as $\ket{\sigma_1^{\dep{k}},\dots,\sigma_8^{\dep{k}}}$, \AFadd{where} $\sigma_i^{\dep{k}}$ is the eigenvalue of $\hat S^z$ on the $i^{th}$ magnetic site for the $k^{th}$ spin configuration, our approach corresponds to identifying $E_{DFT}^{(k)}$ with $\bra{\sigma_1^{\dep{k}},\dots,\sigma_8^{\dep{k}}}  \hat{H}_{ex} \ket{\sigma_1^{\dep{k}},\dots,\sigma_8^{\dep{k}}}$.
        %
        %
        When a sufficiently large number of states is considered (typical pools of states include FM, AFM and some intermediate configurations), the effective exchange parameters of $\hat H_{ex}$ can be extracted from the best fit of the DFT total energies. 
    \begin{figure}
        \centering
        \includegraphics[scale=0.3]{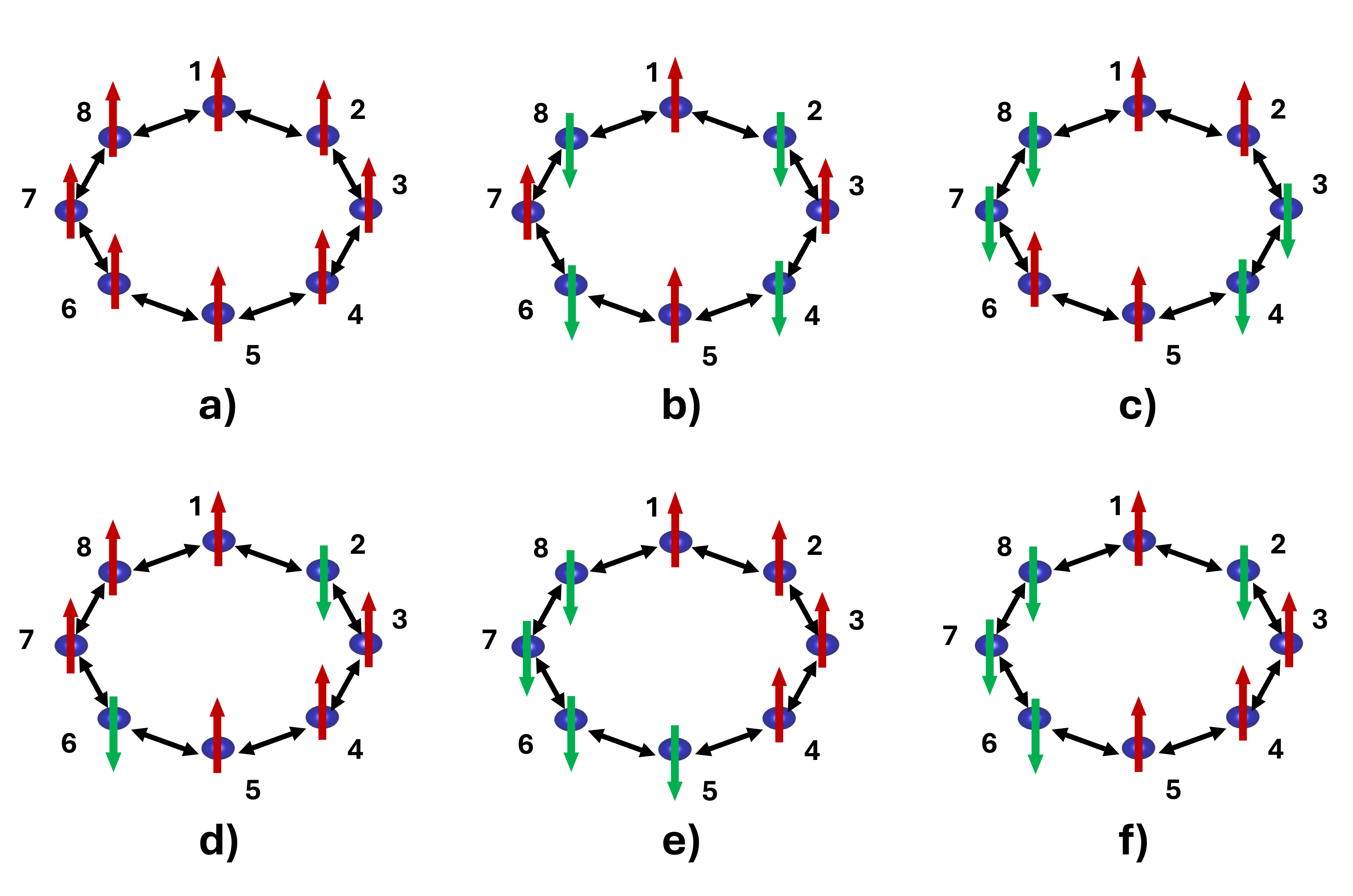}
        \caption{Non-equivalent magnetic configurations used to extract the exchange coupling parameters. a) and b) are the FM and AFM configurations, respectively.}
        \label{fig:j-config6}
    \end{figure}
    Fig.~\ref{fig:j-config6} shows some of the spin configurations typically used in the calculation of the exchange parameters.
    
    After having computed the magnetic couplings, it is worth stressing that,
    while the structure of the molecule possesses a D$_4$ symmetry, \AFadd{due} to the alternate modulation of ligands groups, the magnetic centers are found to be approximately equivalent and deviations from a D$_8$ symmetry can be safely neglected both for the modulus of the localized spins and for the interactions between them. 
    In addition, due to the annular geometry of the systems considered here, periodic boundary conditions are assumed on the site indexes: $i + 8 = i$.

        %
        %
        

%% file: 4-computational.tex
\subsection{Computational details}\label{sec:computational}
 \gls{DFT} calculations were performed using the publicly available, plane-wave pseudopotential package Quantum ESPRESSO (QE) \cite{QE2009,QE2017,QE2020}. 
Spin-resolved, Hubbard-corrected calculations were based on 
a collinear-spin formulation of DFT. Specifically, we compare results obtained from two most popular approximations for the exchange-correlation (xc) functional:
the local density approximation (LDA) \cite{PhysRev.136.B864,KS65}
and the PBE flavor of the generalized-gradient approximation (GGA) \cite{Perdew:1996}.
%
%
The pseudopotentials of the TMs were obtained from the QE PSlibrary archive \cite{PhysRevB.82.075116,DALCORSO2014337} and were chosen to be Ultrasoft (US) and projected-augmented wave (PAW) for GGA and LDA, respectively. 
\gls{KS} wavefunctions and charge density were  expanded in plane-waves up to a kinetic-energy cutoffs of 80Ry and 640Ry, respectively, in all cases.  

\begin{figure}[!t]
    \centering
    \includegraphics[height=5cm]
       {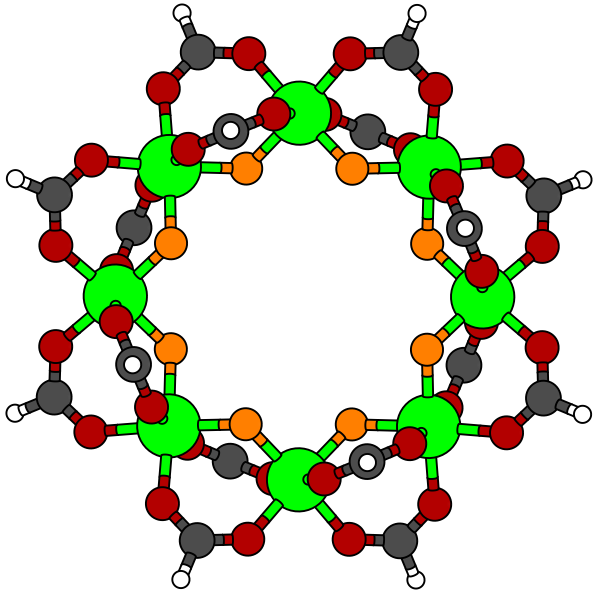}
    \caption{Reduced molecular structure, \AFadd{see} text for explanation.}
    \label{fig:geometries-reduced}
\end{figure}

Due to the complexity of the real  molecular systems, a simplification of the structures was necessary to make the calculations  affordable. This was achieved using the widely adopted ``hydrogen saturation'' procedure \cite{ bellini2010density,bellini2006density, Kronik}, which consists in replacing terminal methyl (-CH$_3$) groups with single H atoms. 
The procedure was sequentially applied twice, and the resulting structure can be observed in Fig.~\ref{fig:geometries-reduced}, \AFadd{to} be compared with Fig.~\ref{fig:geometries-molecule}. %
Overall, the pivalic groups were replaced by -O$_2$CH groups. 
As a result, our simulated systems have chemical formulas Cr$_8$F$_8$(O$_2$CH)$_{16}$ and V$_8$F$_8$(O$_2$CH)$_{16}$. 
%
%
Since the rings magnetic properties stem from the localized \gls{TM} ions $d$-electrons, the reduction is expected to have negligible effects on the systems' magnetic behavior.
Both simplified structures were optimized using both GGA and LDA
and  differences between the  equilibrium structures of Cr$_8$ and V$_8$ were found marginal.
%
Calculations on isolated molecules were performed %
by placing them in empty simulation boxes with a distance between periodic replicas  of  $\approx$ 10 \AA. Correspondingly we used a $\Gamma$-point sampling of the Brillouin zone. 

In order to improve the $d$ states  localization  and achieve a better representation of magnetic properties, we also used  the DFT+U and DFT+U+V schemes.
On-site $U$ and inter-site $V$ Hubbard parameters  were computed 
using linear-response theory \cite{LRHubbard} within the automatic approach introduced in Refs. \cite{Cococcioni2018,timrov21us,timrov22,timrov20pulay}.
To speed-up this calculation, a linear 
model of the rings' local structure  was used, i.e. a chain containing only two magnetic centers \AFadd{instead} of eight, with the same octahedral environment of the TM centers as in the whole molecule. 
\begin{figure}[!htb]
    \centering
    \includegraphics[height=4.2cm]
       {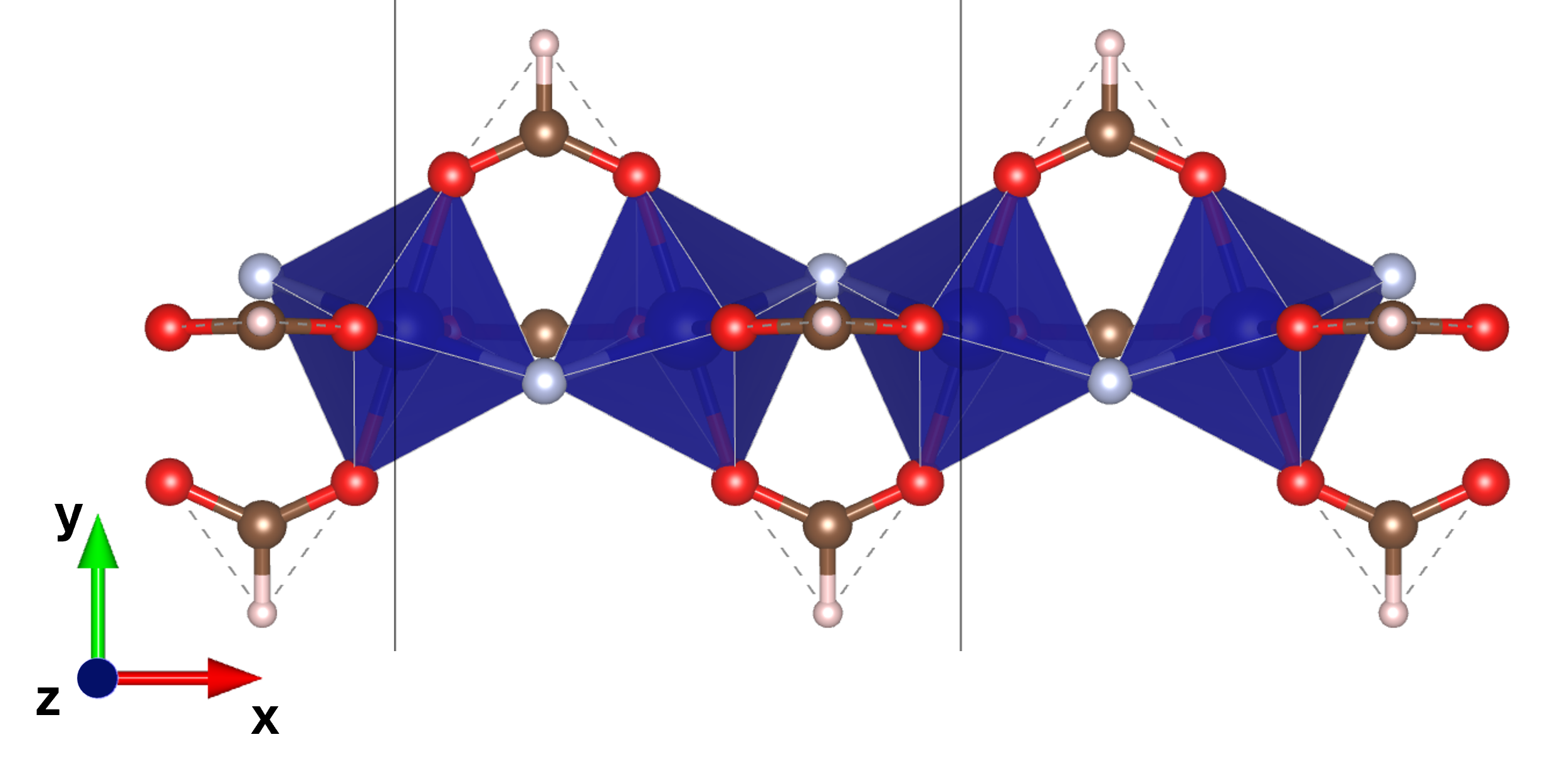}
    \caption{Linearized chain model of the ring molecule, used to compute the Hubbard parameters. The vertical lines delimitate the two-magnetic-sites unit cell, \AFadd{see main} text.}
    \label{fig:chain}
\end{figure}
The chain is obtained from the simplified ring  in Fig.~\ref{fig:geometries-reduced} by changing the position of the ligand groups around the magnetic centers in such a way they can be arranged in a linear conformation. In the process, the structure of the local O-F octahedra around the TMs is maintained as close as possible to the original one.  \AFadd{The main difference between the local structure around the TM centers in the ring and in the chain is in the disposition of the F anions that are not anymore all on the same side towards the internal hole of the ring, but assume alternate positions on either side of the chain axis. Due to this alternate structure the shortest chain that can be used contains two TM units, which is the system used for the calculation of the Hubbard parameters.} Test calculations of the Hubbard parameters in the ring  
and in the chain \AFadd{optimized} with GGA, yielded almost coincident results \footnote{A more exhaustive discussion about this procedure \AFadd{and} the importance of a comparison between curved and linear molecules, will be given in a separate publication}.

In all cases the calculation of these parameters was performed self-consistently with respect to the electronic state: \AFadd{$U$'s} and $V$'s were obtained perturbing a DFT+U(+V) ground state until input and output values converged.
%
The obtained values are 
reported in Table \ref{table:Hubbard-parameters}, showing, 
for both systems,  the considered xc functionals and flavours of the Hubbard correction (DFT+U and DFT+U+V), the value of the computed Hubbard parameters.

\begin{table}[h]
\centering
    \begin{tabular}{c|c|c r|c r }
       \hline \hline
      & &\multicolumn{2}{c|}{Cr$_8$}  	&\multicolumn{2}{c}{V$_8$}  \\
      		& & DFT+$U$ & DFT+$U + V$ 	& DFT+$U$ & DFT+$U + V$ \\
        \hline
      \multirow{3}{3em}{LDA}&
      \multirow{1}{1em}{$U$} & 5.26	& Cr-Cr 6.28	& 4.62	& V-V 5.66	\\
                        & \multirow{2}{1em}{$V$}
                        & & Cr-F 1.54	& 	& V-F 1.48	\\
            			&	& & Cr-O 1.40	&	& V-O 1.37  \\
        \hline
      \multirow{3}{3em}{GGA} & 
      \multirow{1}{1em}{$U$} & 5.13	& Cr-Cr 6.11	& 4.51	& V-V 5.53	\\
            			& \multirow{2}{1em}{$V$} & & Cr-F 1.48	& 	& V-F 1.45	\\
            			& &	& Cr-O 1.37	&	& V-O 1.37	\\	
        \hline \hline
    \end{tabular}
\caption{Calculated Hubbard parameters (in eV).}
\label{table:Hubbard-parameters}
\end{table}
%
It is important to remark that the results shown in Table \ref{table:Hubbard-parameters} refer to the ground-state spin configuration of each system.
%
However, test calculations have shown that the Hubbard parameters' dependence on the arrangement of magnetic moments can be safely neglected.
Table \ref{table:Hubbard-parameters} shows
relatively minor differences (0.1 - 0.2 eV) in the value of $U$ and $V$ between LDA and GGA, independently of the flavor of the Hubbard correction. In contrast, turning on the inter-site interaction ($V$)  leads to a significant upward shift ($\approx$ 1 eV) of the on-site Hubbard $U$. At the same time the intersite $V$ exhibits surprisingly similar values for Cr and V, probably due to the close resemblance of the ligands' structure around the TMs in the two systems.
The on-site Hubbard parameters show instead a more pronounced difference
between the two systems with those of Cr  $\approx$ 0.6 eV higher than those of V.
Finally,  LDA yields higher values than GGA; the shift, \AFadd{which is} non uniform and slightly more pronounced for  $U$ than for $V$, is of the order of 0.1 eV.

%% file: 5-results.tex
    \section{Results}
    \label{sec:res}
        
        %
        

        %
%

        The main objective of this work is to evaluate the superexchange  couplings between the magnetic sites of the considered molecules.  Since  we focus on 
        collinear-spin calculations, only the isotropic part of the effective interactions are obtained \footnote{Higher-order terms will be evaluated and discussed in a future publication, also employing a non-collinear-spin DFT}.       %
%
        %
        %
        %
        %
        Within these settings the expectation values of $\hat H_{ex}$ on the collinear spin states corresponding to the ones computed with DFT reduce to the energy of a classical Ising model: 
        %
        %
        \begin{align}
            E^{\dep{k}} = & \,
            \Jone \sum_{i} \sigma_i^{\dep{k}}\sigma_{i+1}^{\dep{k}} 
            + \, J^{\dep{2}} \sum_{i} \sigma_i^{\dep{k}}\sigma_{i+2}^{\dep{k}} + c,
            \label{energy-function-2}
        \end{align}%
        where the additive constant $c$ fixes the energy reference.
        The considered spin configurations are identified by the sign of the spin on each magnetic site (the modulus being always maximum: $\vert \sigma_i \vert = S_i$) which is controlled by the proper initialization of the local magnetization. Typically, between the FM and the AFM states  we consider five/six intermediate configurations, \AFadd{some}  exemplified in Fig.~\ref {fig:j-config6}, featuring various numbers of spin-flips in various positions. 
        
        A first rough estimate of the nearest-neighbor  couplings can be also obtained from the direct comparison of the FM and AFM spin arrangements, \AFadd{see} Fig.~\ref {fig:j-config6}. The results  for LDA and GGA are reported in Table \ref{ggalda}, showing 
            \begin{table}[h!] 
            \begin{tabular}{C{1cm}|C{2cm}C{2cm}}
                \hline \hline
                &Cr$_8$     &V$_8$   \\
                \hline
                ${\rm J}_{LDA}$  &  16.043    & -38.370 \\ 
                ${\rm J}_{GGA}$ &  3.498     & -3.434      \\
                \hline \hline
                \end{tabular}
            \caption{\label{ggalda}Nearest-neighbor exchange  parameters (in meV) extracted from FM and AFM states.}
           \end{table}
         that both functionals yield quite large exchange couplings~\footnote{We stress that results in Table \ref{ggalda} are rather sensitive to the molecular structure (they change significantly for minor atomic readjustments) which, in turn, is hard to optimize as the convergence at the electronic level is often difficult, especially for the LDA Cr$_8$ value, not fully converged. This sensitivity disappears when using DFT+U or DFT+U+V (Tables \ref{table:J-fit-collinear-ggalda+u} and \ref{table:J-fit-collinear-ggalda+uv}). We thus believe it to be another effect of the overdelocalization of valence $d$ electrons in LDA or GGA.}. In the case of Cr$_8$, the values strongly overestimate the experimental one, $\sim$1.46 meV, from inelastic neutron scattering (INS) data \cite{PhysRevB.67.094405} and electron paramagnetic resonance (EPR) measurements \cite{van2002magnetic,Waldmann03}. The mismatch  is  larger in LDA, which for V$_8$ yields an unrealistic  $\sim$-38 
         meV value.
        Besides 
        these quantitative inaccuracies, it is important to notice the difference in the sign of the  couplings between V$_8$ and Cr$_8$. 
        In agreement with literature \cite{baker2012spin,van2002magnetic, Waldmann03,PhysRevB.67.094405, Kronik, slusarski2011electronic, bellini2010density, bellini2006density, Bellini2008}  Cr$_8$ is found to be AFM,  driven by positive nearest-neighbor exchange interactions $J^{(1)}$. 
        In contrast, a completely different picture emerges for V$_8$, exhibiting
         a negative \Jone that
        stabilizes a FM order. 
        

        Importantly, using Hubbard corrections strongly modifies the exchange interactions. In the following we present the values obtained with DFT+U and DFT+U+V \AFadd{for both systems, where} DFT is either LDA or GGA. In this case the fit on DFT total energy was based on a larger number of spin configurations (see Fig.~\ref {fig:j-config6} for a selection) which also gave access to the exchange couplings between second nearest neighbors. \AFadd{Those} between further neighbors were found negligible in all  cases.
        \begin{table}[!t]
            \begin{tabular}{C{2cm}|C{0.8cm} |C{2cm}C{2cm}}
                \hline\hline
                Functional & $J$ & Cr$_8$ & V$_8$\\
                \hline
                \multirow{2}{*}{LDA+U}& \Jone & 0.843 & -0.917 \\
                & \Jtwo & -0.013 & 0.231 \\
                \hline
                \multirow{2}{*}{GGA+U}& \Jone & 0.628 & -0.937 \\
                & \Jtwo & -0.011 & 0.098 \\
                \hline \hline
            \end{tabular}
            \caption{Cr$_8$ and V$_8$ exchange couplings (in meV) from collinear LDA+U and GGA+U calculations. 
            }
            \label{table:J-fit-collinear-ggalda+u}
        \end{table}
        Table \ref{table:J-fit-collinear-ggalda+u} collects  the couplings obtained from LDA+U and GGA+U, \AFadd{with} on-site corrections only.
        While  the couplings upon the Hubbard correction are hugely renormalized as compared to LDA and GGA, the signs of the nearest neighbor couplings are  not affected, meaning that the character of the ground state in both systems remains the same: AFM for Cr$_8$ (\Jone $>$ 0) and FM for V$_8$ (\Jone $<$ 0). The values of \Jone  
        are now of the expected order of magnitude and are much closer  to the  
        experimental value for Cr$_8$ ($\sim$1.46 meV \cite{PhysRevB.67.094405,Waldmann03,van2002magnetic}). 
        Both LDA+U and GGA+U show somewhat underestimated values. 
        
        Moreover, besides a FM \Jone, V$_8$ also features a sizeable second-nearest neighbor coupling \Jtwo, that is positive and, in magnitude, 
        represents a significant fraction of \Jone. In fact, a sizeable \Jtwo is needed in order to fit the DFT total energies and  make the system "Heisenberg-like", \AFadd{i.e.}, with  couplings independent from the spin configuration. \Jtwo is also present for Cr$_8$ but is one order of magnitude smaller than for V$_8$.        
        The presence of a relatively large and positive \Jtwo is  an interesting feature of V$_8$:  competing (opposite sign) magnetic interactions of comparable strength might induce  frustration and/or enhance the system's susceptibility to external fields,  enabling easier switching between different magnetic orders, an aspect that could be exploited in some applications. 
        

        In order to fully appreciate the role of the Hubbard corrections we also evaluated the couplings within the generalized DFT+U+V scheme \cite{DFTUV}. The inter-site effective interaction ($V$) between electrons localized on neighboring sites is likely to influence  the  couplings between localized moments, typically mediated by anions' extended orbitals between them. The results are shown  in Table \ref{table:J-fit-collinear-ggalda+uv}.
        \begin{table}[t!]
            \begin{tabular}{C{2cm}|C{0.8cm} |C{2cm}C{2cm}}
                \hline\hline
                Functional & $J$ & Cr$_8$ & V$_8$\\
                \hline
                \multirow{2}{*}{LDA+U+V}& \Jone & 1.199 & -0.577 \\
                & \Jtwo & -0.014 & 0.248 \\
                \hline
                \multirow{2}{*}{GGA+U+V}& \Jone & 0.908 & -0.762 \\
                & \Jtwo & -0.011 & 0.174 \\
                \hline \hline
            \end{tabular}
            \caption{Cr$_8$ and V$_8$ exchange couplings (in meV) from collinear LDA+U+V and GGA+U+V calculations. 
            }
            \label{table:J-fit-collinear-ggalda+uv}
        \end{table}
        The comparison with Table \ref{table:J-fit-collinear-ggalda+u} results clarifies immediately that the  coupling $V$ \AFadd{between} the TMs and the anions in their first shell of neighbors, generally shifts the value of  couplings upward, \AFadd{with} negative ones \AFadd{becoming} less negative. This slightly improves the agreement for Cr$_8$ with experiments compared to the DFT+U case. The shift is quantitatively more consistent for \Jone than \Jtwo and slightly more pronounced for LDA than  GGA. Aside from these differences, however, the overall picture remains unchanged: positive (AFM) \Jone's and negligible \Jtwo for Cr$_8$; negative (FM) \Jone and consistent, positive \Jtwo for V$_8$. 

        \AFadd{In spite of their structural and chemical similarity, as Cr and V are adjacent on the periodic table, the two molecules exhibit radically different ground states: Cr$_8$ shows an AFM order mainly stabilized by \Jone, while V$_8$ displays a FM arrangement resulting from the competition between \Jone, which is FM and numerically prevalent, and \Jtwo, which is AFM.}

%% file: 6-discussion.tex
\section{Discussion}
\label{sec:disc}
The results from the previous Section  carry several messages that we can now examine in some detail.

First, as well known, LDA and GGA functionals are inadequate for accurately capturing the d-states localization. 
The consequences of this inaccuracy are  dramatic for the properties studied in this work. In fact, given the typical small values of the interatomic exchange parameters, even a slight electron delocalization might lead to a gross overestimate of the couplings, as shown in Table \ref{ggalda}.
This traces back to the physics of superexchange, which mediates magnetic interactions between localized moments via intermediate anions with fully occupied orbitals. In fact, in a simple Hubbard model of the electrons responsible for the localized moments, the magnetic coupling, to the lowest order of perturbation theory, is proportional to $t^2/U$. Here $t$ 
represents the hopping parameter (or bandwidth), which quantifies the energetic advantage of forming bonds or crystalline itinerant states, 
while $U$ is the short-range on-site Coulomb repulsion. 
Clearly, overestimating 
$t$ and underestimating 
$U$, both consequences of the local/semilocal functionals, leads to severely off-scale magnetic interactions. This interpretation also explains the significant improvement in evaluating the exchange couplings achieved with DFT+U, which incorporates corrections to local/semilocal functionals inspired by the Hubbard model. By using LDA+U and GGA+U, we were able to predict \Jone for Cr$_8$ with much closer agreement to the experimental value compared to LDA and GGA and, in general, obtain exchange couplings of the correct order of magnitude.

The agreement with experimental results on Cr$_8$'s \Jone is further enhanced by using the extended DFT+U+V scheme with on-site and inter-site interactions \AFadd{between} TM magnetic centers and neighbor anions. While this result probably confirms the role of the intermediate anions orbitals in the superexchange mechanism,
the overall qualitative picture is the same as with DFT+U.

Regarding the comparison between LDA and GGA, based on the  value of \Jone in Cr$_8$, LDA produces slightly more accurate results than GGA when corrected with some flavor of the Hubbard functional. While this result does not include the effects of structural relaxation, we can say that, again, the overall qualitative picture emerging from the two families of functionals is the same.
However, in the perspective of performing non-collinear spin calculations, GGA xc kernels are sometimes reported to be more problematic than LDA \cite{pu22}. Thus, LDA might be slightly preferable, especially when using a basic non-collinear-spin implementation of DFT.  

From a physical point of view, the most important outcome to remark is  the different behavior of Cr$_8$ and V$_8$.
In spite of the identical ligand struture and of their chemical similarity, independently also from the specific functional used to model their behavior, the molecules exhibit a radically different ground-state magnetic order: AFM for Cr$_8$, FM for V$_8$. In addition, while the energetics of Cr$_8$ can be adequately described using only the nearest-neighbor exchange interaction \Jone, V$_8$ also exhibits a significant next-nearest-neighbor coupling, \Jtwo, with a magnitude comparable to \Jone but of opposite sign.

This behavior of V$_8$ is rather unusual. Our preliminary calculations, where V centers are substituted with other TMs, consistently  
stabilize an AFM ground state.
The V$_8$ peculiarity  
seems to question the validity of the superexchange theory (also challenged by the presence of multiple anionic links between neighbor TM centers) and of the Goodenough-Kanamori-Anderson rules. However, it is important to note that in their simpler form, superexchange mechanisms are often referred to half-filling conditions. These mechanisms are  derived
from second-order perturbation theory applied to an Hubbard model with one orbital and one electron per site (half-filling).
This condition is not satisfied by V$_8$ that, in its 3+ oxidation state, hosts the two outermost electrons in a group of three quasi-degenerate $d$ states, \AFadd{namely} the $t_{2g}$ multiplet if the ligands' 
octahedron around it had a perfect cubic symmetry. We believe that this less-than-half-filling condition of  V$^{3+}$  
\AFadd{is at the origin of V$_8$’s FM ground state. In fact,}
it is quite well established in literature that at less-than-half filling, a strong enough on-site repulsion between electrons ($U$ or $U-J$ for parallel spins) or, equivalently, a weak hopping amplitude ($t$), can cause a FM coupling between neighboring spins, mediated by the intermediate anion  
\cite{schulenburg97,weber98}. 

\begin{figure}[!t]
    \centering
    \includegraphics[height=6cm]{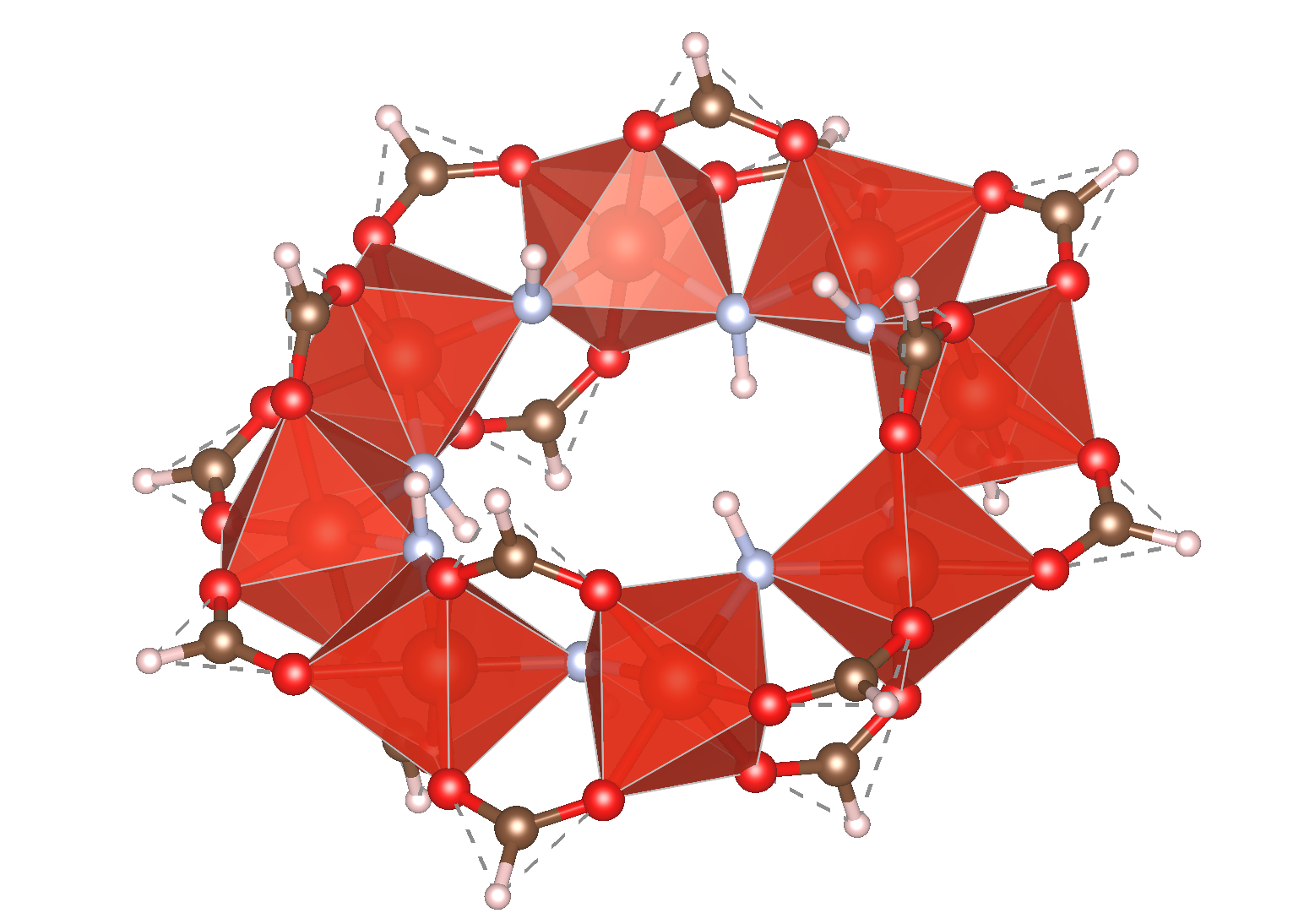}
    \caption{A modified V$_8$ ring highlighting the presence of one extra H bound to each F atom in the internal space of the molecule.}
    \label{fig:FH}
\end{figure}
\AFadd{In order to verify this idea, we have performed a series of  tests on modified V$_8$ or Cr$_8$ rings that 
feature a different anionic group at the F site: either an H-coordinated F or an O, so as to change the number of $d$ electrons on the magnetic centers. Fig. \ref{fig:FH} shows, as an example, the ring with an extra H bound to each internal F.  
It is important to note that besides the modification of the anionic center, 
no extra geometrical optimization
has been performed. The evaluation of the effective 
$J$'s was achieved by comparing the 
FM and AFM configurations energies, computed with LDA+U.} 
        \begin{table}[h!]
            \begin{tabular}{C{1.5cm}|C{1.5cm} |C{2.2cm}C{1.5cm}C{1.cm}}
                \hline\hline
                System & Anion & Av. mag. ($\mu_B$) & $J$ (meV) & stable \\
                \hline
                \multirow{2}{*}{Cr$_8$}& F & 3& 0.843 & AFM \\
                & O & 2  & -148.188 & FM \\
                \hline
                \multirow{3}{*}{V$_8$}& F & 2 & -0.917 & FM \\
                & F-H & 3 &  7.576 & AFM\\
                & O & 1 &  -22.456 & FM\\
                \hline \hline
            \end{tabular}
            \caption{\AFadd{3$d$ states occupations and effective $J$ for the modified V$_8$ and Cr$_8$ systems (see text).} 
            }
            \label{table:modr8}
        \end{table}

\AFadd{
Table \ref{table:modr8}  compares the average absolute magnetization per TM center and the $J$'s 
for Cr$_8$ with F replaced by O, and 
for V$_8$ with F replaced by F-H and O. ``Standard" values, obtained with F on the internal anionic site are also reported to ease the comparison (these are the same as in Table \ref{table:J-fit-collinear-ggalda+u}). 
}
\AFadd{In all cases the substitution of F anions with F-H groups (with single O) leads to a reduction (oxidation) of the TM centers that couple to a consistent unitary increase (decrease) in the absolute magnetization of the TMs. 
The V centers reduction 
brings 3 electrons to the $3d$ states 
(like Cr in Cr$_8$) which achieves a 2+ state and a half filling condition for the occupied submanifold. Correspondingly, the 
$J$ becomes positive and large 
(7.576 meV) which stabilizes an 
AFM configuration. 
On the contrary, with O replacing F, the TM centers get 
a 4+ state that promotes a further departure from half-filling. With less than half-filled $3d$ submanifolds both systems exhibit negative $J$'s 
that make FM spin configurations more stable. In this case
the $J$'s are 
very large in absolute value (-22.456 meV and -148.188 meV for V$_8$ and Cr$_8$, respectively). 
Due to the high energy required to reach a 4+ oxidation state 
the hole introduced in the structure 
is actually spread over the rings with a consequent severe overestimation of the computed couplings (similarly to what is obtained with LDA or GGA).
However, the main focus of this discussion is the sign of the magnetic couplings.
Our conjecture is thus overall confirmed: 
magnetic centers with 
less-than-half-filled sub-groups of $3d$ states can couple ferromagnetically, even for geometries that would otherwise promote AFM configurations.  Therefore, the anomalous ground state of V$_8$ can, at least in part, be traced back to the lower filling of its $d$ states.}

%% file: 7-conclusions.tex
\section{Conclusions}\label{sec:conclusions}
In this work we have presented a computational study of two ring-shaped molecular nano-magnets with eight magnetic sites, named Cr$_8$ and V$_8$. Fitting the DFT energies corresponding to various collinear-spin configurations we were able to extract the inter-site exchange magnetic couplings between the localized spins, using a number of different exchange-correlation functionals and Hubbard-based corrective schemes. 
Consistent with previous studies, we find that standard LDA/GGA DFT approximations fail to localize valence electrons on the $d$ states of the transition-metal magnetic centers, resulting in gross overestimations of the magnetic couplings. 
Energy corrections introduced with DFT+U and DFT+U+V schemes lead to significant improvements,  with the inter-site coupling $V$ further refining the agreement with the experimental measurements.
Independently of the functional used, and in agreement with previous investigations, Cr$_8$ is found to have an antiferromagnetic ground state, with magnetic interactions limited to nearest-neighbors.
On the other hand, V$_8$ is found to have a ferromagnetic ground state with nearest-neighbor interactions competing with antiferromagnetic next-nearest-neighbor interactions of the same order of magnitude.

The anomalous behavior of V$_8$ can be 
\AFadd{traced back to the} less-than-half filling condition of the $d$ states submanifold of V$^{3+}$ hosting the valence electrons responsible for the ion's localized moments. \AFadd{This idea has been proved by evaluating the effective exchange coupling in modified systems where the number of $d$ electrons on the TM ions was artificially changed by either coordinating F ions with extra H atoms or substituting them with O anions.}

%% file: acknowledgements.tex
\section{Acknowledgements}

\AFadd{The authors are thankful to Prof. L. Kronik for a preliminary discussion on the Cr$_8$ system and for providing its initial structure.} B. M. M. and M. C. acknowledge  support from  NQSTI within the PNRR MUR project PE0000023-NQSTI.
Part of the calculations of this work were realized thanks to the high-performance computing resources and support made available by CINECA (through awards within the ISCRA initiative). 
Moreover, via A.F. membership of the UK’s HEC Materials Chemistry Consortium, funded by EPSRC (EP/X035859), this work also used the
ARCHER2 UK National Supercomputing Service (http://www.archer2.ac.uk).